# Break-in' Point: Somatic Narratives - the convergence of arts and science in the transformation of temporal communities

Carol Marie Webster[1], Panagiotis Pantidos[2], DeNapoli Clarke[3] and Jiannis K. Pachos[4]

[1]Independent Researcher, currently based in New York City, U.S.

[2]Department of Early Childhood Education, National and Kapodistrian University of Athens, Greece

[3]Northern School of Contemporary Dance, Leeds, LS7 4BH, UK

[4]School of Physics and Astronomy, University of Leeds, Leeds LS9 2JT, UK

## Abstract

*Break-in 'Point,* a 2012 arts and science performance and community engagement research initiative, was presented in the spring and fall semesters at the University of Leeds in the UK at Stage@Leeds. The outcome of a collaboration between dance artist A3 and theoretical physicist A2, under the direction of performance researcher A1, *Break-in 'Point* is based on a series of real-life encounters at intersections of arts and science - exploring force, risk, exposure, and resilience. The *Break-in 'Point* performance offered an interrogation of the critical point at which physical, mental, and/or emotional strength give way under stress - causing structural degeneration and the experience of what lies beyond. This article is an examination of the performance, reviewing and analyzing it as an imagined somatic zone [1] that engaged spiritual and epistemological transformation of performers and audiences. The article addresses three main periods in the life of *Break-in' Point*: 1) the development period - script building and rehearsals; 2) the performance - live encounters between and among performers and audiences; and 3) beyond the theatre - digital engagements in the classroom and pedagogy. The article contributes new concepts and new ways of thinking about science education, the role of digital technology in pedagogy, dance/theatre public engagement, and community arts practices as practices of healing, health, and resilience.

## Introduction

Every human experimentation begins with driving curiosity, a willingness to take certain calculated and un-calculated risks and a hopefulness on their outcomes: "The nature of somatic work is experiential. It takes the [practitioner] beyond discernment of the body that is only intellectual towards an embodied or experienced understanding" (Geber and Wilson 2010, p. 52). With embodied (or bodily experienced) understanding as a valued site of knowledge the experiential nature of somatic work takes on an enlivened epistemological stance whereby somatic work is intellectual and the body is simultaneously the articulated and articulating site of intelligence and intellectual inquiry (Geertz 1973, De Certeau 1984). With this, it is understood that human experimentation is experiential: it is risky business - with the body as the primary access to knowledge, it is emotional and it is sensorial, it is rich and thick with humanity (Harrison, Walker and Edwards 2002, Lepecki 2006, Profeta 2015).

---

[1] The term imagined somatic zone points to areas in which embodied encounters transcend temporal bound-ness compelling new possibilities.

In 2012, the arts and science performance *Break-in' Point* was presented as a part of the spring and fall production schedules at Stage@Leeds (University of Leeds, UK). A collaboration between dance artist A3, theoretical physicist A2 (the two performers) and performance researcher A1 (the director), *Break-in' Point* was billed as a series of real life encounters that explore the relation of movement and science, analyzing the critical point at which physical, mental and emotional strength give way under stress causing structural degeneration and mapped the experience of what lies beyond. *Break-in' Point* engaged human trauma and vulnerability at the meeting of arts and science, accommodating necessary slippage between the two in order to fully grasp the complexity of human experience and response.

This article presents an examination of *Break-in' Point,* attentive to spiritual and epistemological transformation of performers and audiences. The concepts *somatic ethics* and *affect* are key analytical tools applied to this reflection on three periods in the life of the project: 1) the development period, which involves script building and rehearsals; 2) the performance that includes live encounters between and among performers and audiences; and 3) beyond the theatre that explores pedagogical applications of the performance in the classroom by employing its digital recording, which can be accessed at https://www.youtube.com/watch?reload=9&v=wqBWcHUjLeU. In each period (representing process, product, and dissemination, respectively), temporal communities were constructed. With attention to these three periods, the authors offer varying perspectives on body episteme (bodily knowledge and bodily knowing) in critical exploration of teaching, learning, and pedagogy; embodied narrative engagement between arts and science; and deep encounters of temporal communities. Temporal communities identify communities formed through situational agreement to varying types of collective participation bound by the limits of time; examples include bus and train riders, dance and theater audiences, and the classroom. The article proceeds in the following manner: starting with where the *Break-in 'Point* journey began, A2 and A3 discuss the process of script building in their reflection on embodied experiential methods developed in the rehearsals and applied in the performances; A2 and A3 highlight the journey of colliding narratives in the communication between body episteme. Then, A1 offers an examination of the affective and somatically charged temporal communities ignited between and among performers and audiences in real and imagined somatic zones. Finally, A4, a physicist and teacher educator who joined the *Break-in' Point* team through its digital form, presents an exploration of encounters with *Break in' Point* beyond the live theatre. This examination addresses ways in which digital engagements in the classroom can inform somatic pedagogy, which is itself a process of meaning-making.

## Guiding terms

The terms *somatic* and *ethics* are brought together to inspire ways of thinking through embodiment as essentially an ethical endeavor. *Somatic* points to the full sensorial experience the body can offer at any given moment in time, and is understood as vast collections of bodily information and knowledge. *Ethics*, most simply refers to a set of principles that inform thoughts and actions. *Somatic ethics* refers to varying ways in which information and knowledge are understood and stored in the body and made use of in the act of daily living. The implicit and explicit ethics visible throughout the performance is what makes *Break-in' Point* stirringly powerful.

In their respective publications, Teresa Brennan (2004) and Sara Ahmed (2004) unpack the concept of affect, asserting that affect is a type of physical materiality that must be reckoned with. Affect is different from other types of materiality in that it compels presence, bodily experience in the moment and relational sensorial engagement with visible and invisible material. Affect is *out there*, in the physical world and is something that can be cultivated by social interactions, while it can be left in a place/space/environment or a zone. Examples include the sense of reverence experienced when entering a religious space, a hospital, or theatre; or, the giddiness that arises on entering an amusement park or a children's birthday party. Affect is also *in here,* meaning inside the body, cultivated by the social and ingested into the inner linings of the corporal. Examples include imposter syndrome, varying forms of shame, and low self-esteem. In Ahmed's view, affect is a substance/material that 'sticks to' the surface of the skin, binds to the body's surfaces, impacting (at times assaulting) the senses as a part of the experience of engaging the external environment, the experience of the externalized zone. For Brennan, affect 'gets into' the body, is invasive, impacting the ways in which people experience spaces/places of the body - their own body and the bodies of others. *Affect* is created in and by the social, and ultimately, serves the social by marking and infecting bodies in encounters within and between real and imagined somatic zones. Zones are spaces of influence and experience. Real zones are spaces occupied in real-time in-person encounters. Imagined zones are those constructed in and between the virtual and the real as well as those constructed in the imagination. For the purposes of this article, all virtual spaces are understood as imagined zones and the construction of encounters between the virtual and real are also understood as imagined zones. Ahmed (2004) and Brennan (2004) point out that affect targets bodies and concerns bodies in relation.

In line with Ahmed (2004) and Brennan (2004), in their examination of dance education practices, Geber and Wilson (2010, p. 53) point to the important relationship between somatics and pedagogy, noting that "as a methodological approach, somatics emphasizes the process of learning […], helps students overcome physical obstacles they may be dealing with in their bodies, and celebrates the "distinct humanness and wholeness of each student." Geber and Wilson (2010) write from the perspective of learning encounters that take place in the dance class/dance studio, yet their theory is not limited to the bodily art of dance or bound by the mirrors and doors of the dance studio. Experiential encounters in and between bodies in the conventional classroom can also allow knowledge to be gathered, assessed, and valued in ways that deepens and enhance bodily knowing, and speaks to "a pedagogic shift, one that encourages curiosity and agency" (Geber and Wilson, 2010 p. 53). The authors of this article have found resonance in Geber and Wilson (2010) proposal and have extended their statement to include virtual bodies.

In this second decade of the early twenty-first century, classrooms are spaces in which body value and knowledge are regularly understood as in contest with the virtual (digital media and attendant technologies); whereby the virtual is anticipated to disrupt (and be disruptive to) processes of learning, knowledge gathering, and knowledge exchange. Whilst, the *Break-in' Point* performance and experimentation, along with the foray into virtual pedagogy, took place in 2012 and 2014, respectively, the authors have found profound relevance in the current COVID 19 pandemic and whatever lies beyond. Our interpretation of Geber and Wilson's notion takes bodily learning beyond the dance studio and to the performance/theatre/community space, and into the virtual.

## Building the script: An Embodied Experiential Process

In this section, A2 and A3 reflect on the embodied experiential process in rehearsal and in performance, highlighting their journey of colliding narratives in the unpacking of body episteme.

*Break-in' Point* started accidentally. A2, a theoretical physicist, wanted to bring dance artist A3 and his dance troupe to the School of Physics and Astronomy at the University of Leeds to perform for undergraduate physics students. In the fall of 2011, after a short discussion with A1 (Performance Researcher and Project Director), the initial idea evolved into the building of a performance with both A2 and A3 on stage and no dance troupe. The subsequent rehearsal process centered on experimentation as A1, A2, and A3 delved into researching the connection and dissonance between human emotions and physics. The rehearsals brought out rich personal narratives told through science and human emotions, and a performance work was created that spoke authentically about the development process and the outcomes.

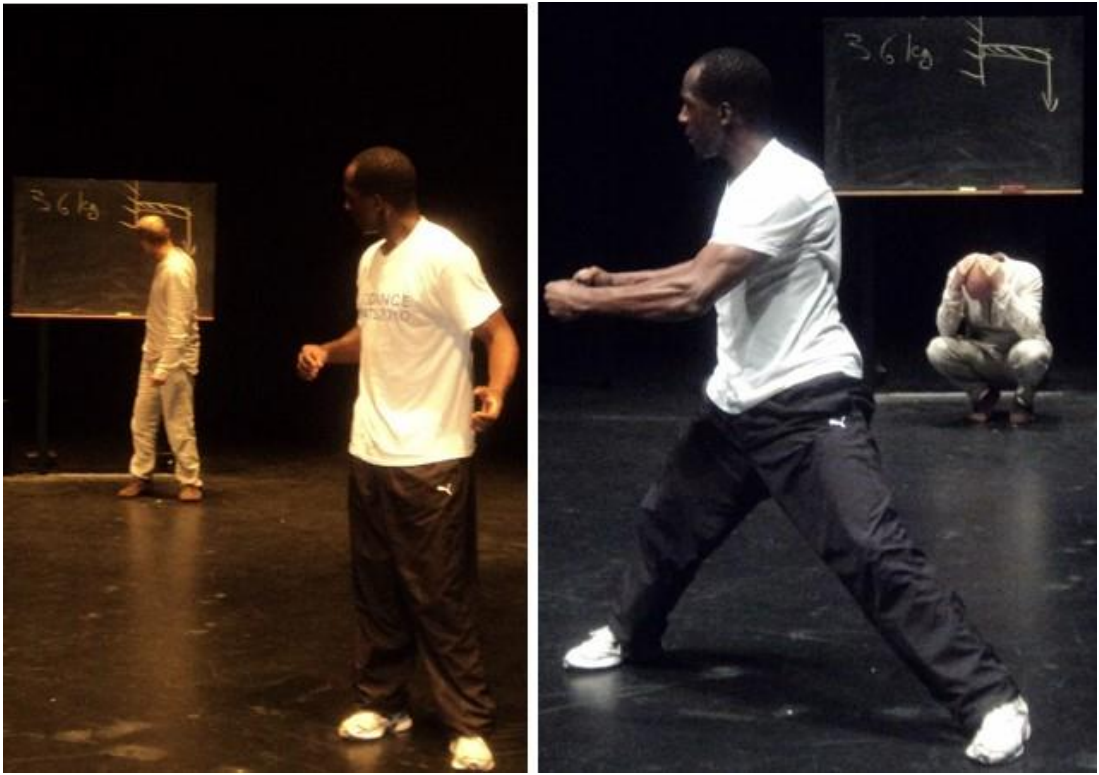

*Figure 1 A2 and A3 in performance at the Stage@Leeds in 2012.*

*Break-in' Point* was from the beginning *experimental*, a research *workshop*. The researchers worked to familiarize themselves with important concepts of physics and dance, engaging through personal and professional narratives. As collaborators from conception to the development of Break-in' Point, three of the researchers, A1, A2 and A3, mutually shared ideas about the process, discoveries, and the personal stories that would support the project's public performance.

## Inside Out Narratives

In the initial instance, there was a central role given to *Halo*, an earlier dance conceived and performed by A3. *Halo* dealt with an incident/attack in A3's life history that resulted in A3 sustaining a broken neck and his path to healing. During the *Break-in' Point* rehearsals A3 shared additional narrative details of circumstances that surrounded the breaking of his neck, as well as information about the continued impact it had on his life as a dance artist and human being. In the initial stages of the *Break-in' Point* rehearsal, A2 deployed physics concepts to describe certain moments from A3's story. A2 drew language from lectures and tutorials common to the physics environment. This initial "scientific" distance had its own narrative, which was brought to light through the experiential rehearsal process. The weaving together of the varying narratives was achieved through A1's directional vision, resulting in an interdisciplinary collaboration between physics and dance without distinct boundaries. The experiential character of the project valued *process* (rather than aiming for a final performance), which gave the performers (A2 and A3) the energy, strength, and conviction that allowed the project to evolve to its live-stage performance phase.

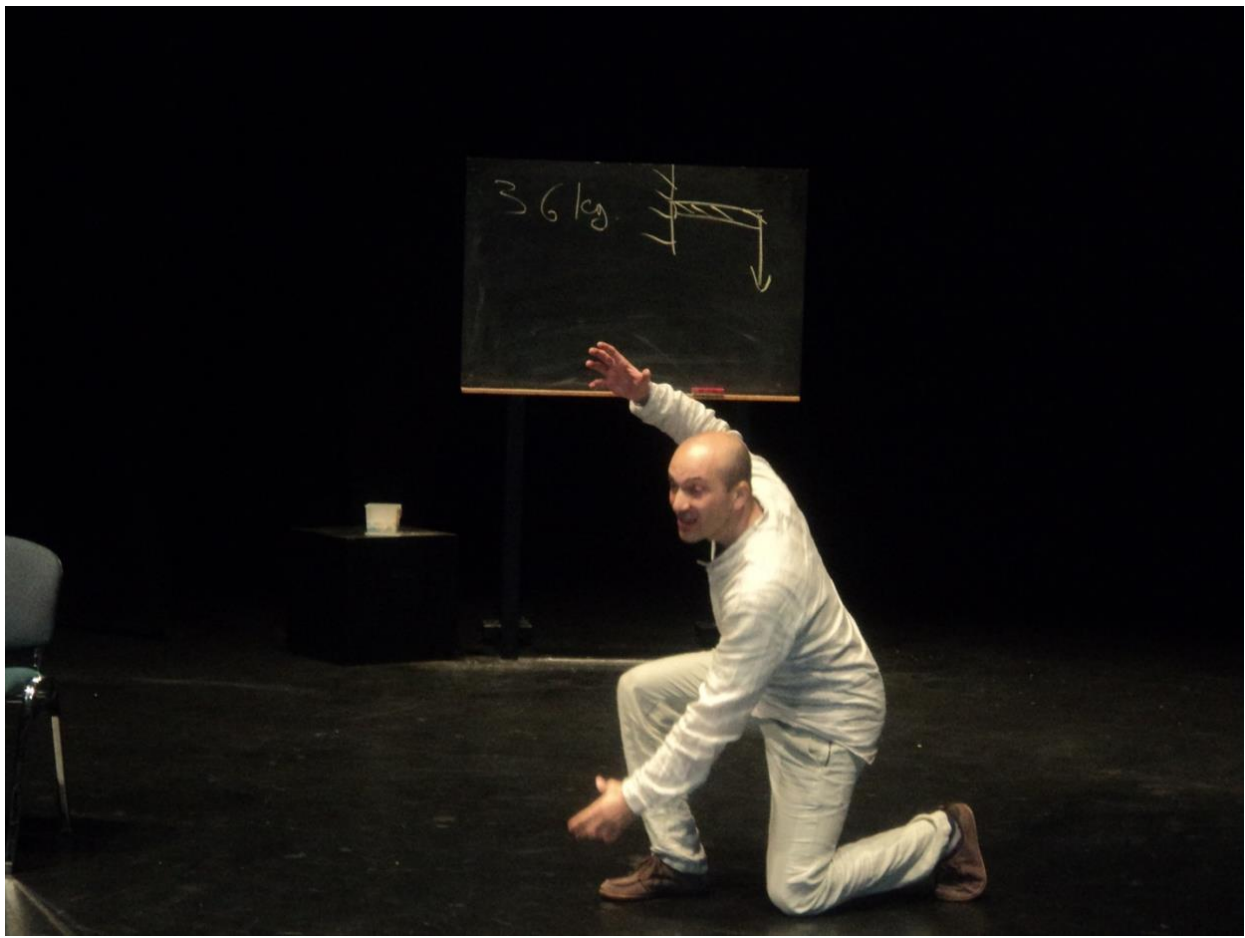

*Figure 2 A2 demonstrates the concept of force during the performance.*

In the rehearsal and performance phases of the project the aim of the physics discussion was to show the human aspect behind laws of physics as well as expose anthropomorphic perception of physics. We called on this as a necessary step to demystify the distance between physical concepts and physical phenomena, and build bridges for ourselves and for non-specialized audiences. The narrative about the breakage of the collar bone was linked with attempts to scientifically study the concept of *force* and the probability for a human being to break their collar bone; this story served as an ideal central narrative bridge.

The concept of force, an essential notion in physics studies, was humanized by A2's narrative of an attempt to teach the concept to a university student who was also a close family member. The question, *What is force*?, elicited a search for a definition that revealed the illusive nature of pinning force to a definition. The description of the failures of a budding scientist to grasp the seemingly simple concept of force, catalyzed A2, a senior scientist and family member, to rethink processes of teaching and learning. In particular, it sparked new perspectives on teaching and learning through personal narratives.

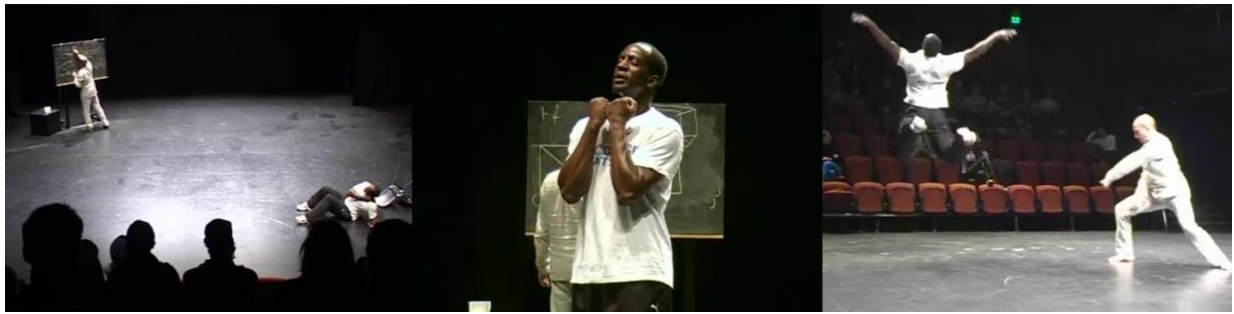

*Figure 3 A3 sharing a personal story of attack and recovery during the performance.*

A3's artistic rendition of the attack that resulted in the breaking of his neck portrayed the collision of physical, emotional, and mental trauma caused by the attack and the circumstances that surrounded it. His recovery process involved a re-imagining and strengthening of his body and mind, which was made possible through the support and contribution of individuals who made up an entire community. In building the script, detailed descriptions were provided of the attack on A3, A3's time in the hospital and the impact on his family, and the manner in which he was strengthened by his family, friends and community. As a result, A3 was able to fully recover, to continue his life as a professional dancer artist and become the director of De-Turf Fitness, a fitness concept based on dance and martial arts movement.

The project title *Break-in' Point* points to the common axis that aligns both physics and dance. The "breaking point" is the critical point at which physical, mental, or emotional strength gives way under stress or pressure. "Break-in" point is also the point in time when a new realization occurs that helps to overcome the seemingly impossible. The alignment of the physics and performance along these two points was facilitated and guided by A1 in an organic way. As we were engaged in the rehearsals we realized a variety of unexpected connections between arts and science. A variety of underlying connections achieved in the performance become apparent to us (A2 and A3) only after we viewed its video recording.

## Discussion: Personal Stories - Narrative collisions and encounters

The performance was scripted, tightly staged and choreographed, the result of the intense embodied experimentation of the rehearsal process. Narratives from the lives of A2 and A3 formed the text/verbal exchange, whilst personal mannerisms, incorporated from the performers' initial sharing of their stories during the rehearsal process, informed the staging and choreography. A2 and A3 shared numerous narratives during the rehearsal process, but only a few made it to the final script; those were the powerful life-altering moments that reverberated and ricocheted through both their lives. The rehearsals were filled with highly charged emotions and angst that eventually resulted in a very personal on-stage experience. Thus, the process required the building of deep trust between the performers (A2 and A3), and between the performers and the director (A1). The rehearsal process required deep levels of emotional commitment that regularly left them physically and emotionally exhausted. As A2 and A3 worked to build resilience, it became clear that resilience needed to be a part of the script. With this, the performers formed a temporal community of courageous survivors who shared their stories as a part of the process necessary for the healing of themselves and of their audiences.

The rehearsals were charged spaces that impacted life outside. Often A2 or A3 reported loss of sleep in their attempts to wrestle and reconcile with the emotional suitcases they unpacked during rehearsal. Although A3 had extensive professional performance experience, in specific with similar types of dance/theatre, this rehearsal process was unsettling. A3 focused his concerns on becoming familiarized with relevant physical concepts and with building the visual and aesthetic temperaments. A3 aimed to "perform as one's self" his understanding of the new physics concepts he was learning (Assaf, 2012). Conversely, A2, a first time performer, had to learn basic performance techniques and needed specific types of support, for this he relied on A3 and A1 at all stages of the project.

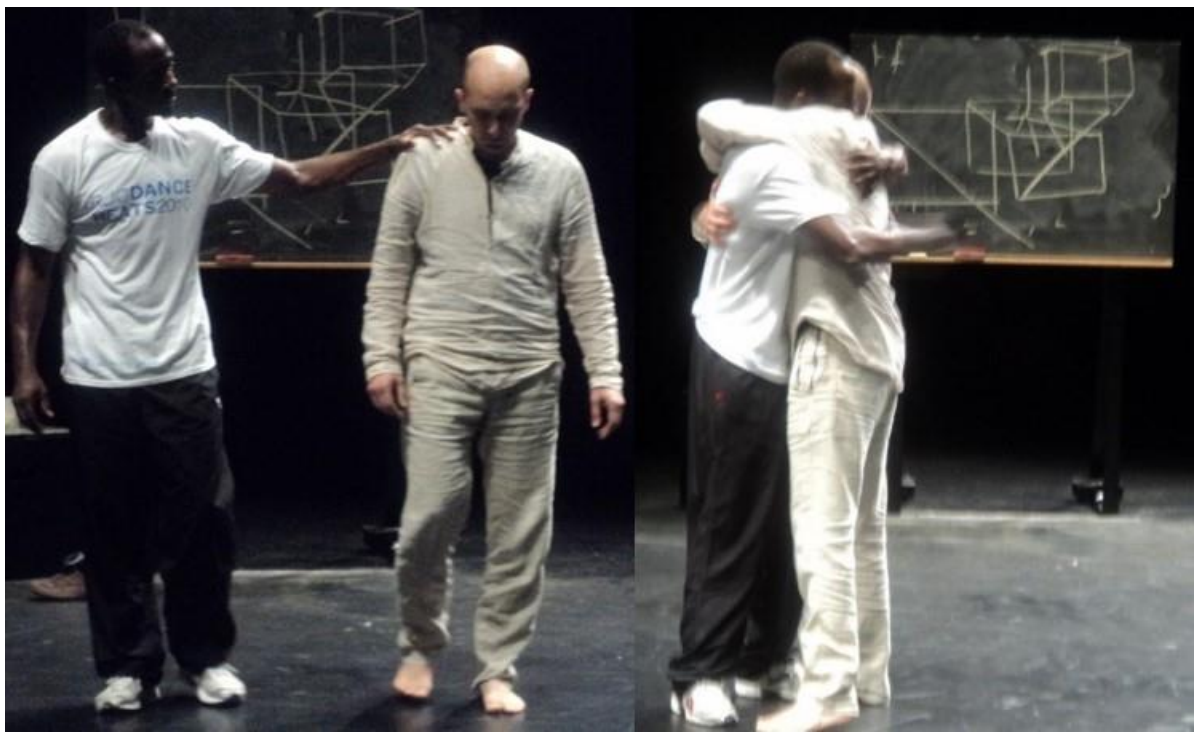

*Figure 4 A3 and A2 comforting each other during the final moments of the 'Break-in' Point' performance.*

Furthermore, during the project we had to create common spaces in which we were comfortable to trust, embrace openness in communication, share stories and experiences, exchange knowledge and learn from each other. Several aspects of the project had to be worked out and communicated, such as the methodologies, physics concepts, performance participation, motivation and the verbal and movement languages. The interdisciplinary nature of the project (the combining of physics and dance) demanded a willingness to take risks, engage in new and unfamiliar embodied knowledge exchanges, and to trust one another as skilled experts in our respective fields. As a team, we were fully committed members of this newly formed temporal community and willingly experienced the power of somatic ethics that kept us honest with ourselves and each other.

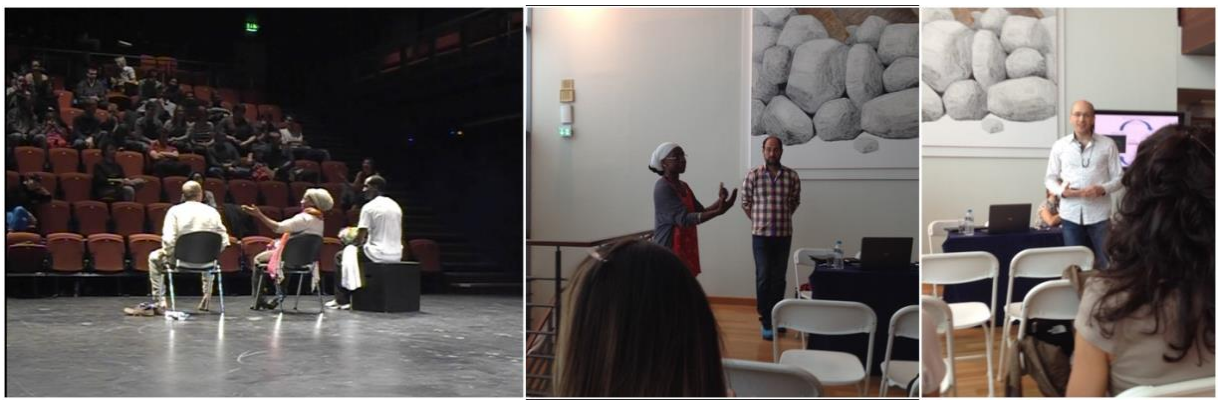

*Figure 5 Break-in' Point audience in conversation with director (A1) and performers (A2 and A3). Right: A1, A4 and A2 presenting their work at the 2015 Society of Dance History Scholar Conference in Athens, Greece.*

As a reminder, somatic ethics is active knowing and knowledge gathering that is tuned to a spirituality that is sensorial; somatic ethics requires bodily engaged reasoning, involves a harmonization of *knowing* that is in and through the body and directed towards justice. Moreover, somatic ethics engages accountability that is committed to truth-telling. Geber and Wilson (2010 p. 52) note that "The focus of the *information transfer is relative to the individual* dancer, *and involves feeling, sensing, imaging and self-identifying* in the context of dancing. *Experiencing the body kinesthetically involves multiple ways of orienting and learning"* (My Italics). Through the collection of information that involves feeling, sensing, imaging, and self-identifying, an individual is involved in varying ways of orienting and learning. With this understanding it can be seen that regardless of the size of the community, somatic ethics is deep-level community building.

## The Risky Business of Embodied Experimentation

> "I didn't expect it to be so distressing… I thought that it would be a nice sweet afternoon performance but it wasn't. . . It was filled with sooo much emotions... I really, really enjoyed it. . ." Audience Member #5 (AM5), 2012.

The *Break-in' Point* project began with the understanding that learning and performance mandate the building of temporal communities and risk is central to these processes. Simply put, risk is in continuous relation with exposure and danger, imagined or real. Exposure is, according to Mariam-Webster Dictionary, "the fact or condition of being exposed… the condition of being presented to view or made known." Exposure is about types of unprotected revealing that presumes the presence of others, and/or being the subject of others' concerns. Exposure immediately points the way to the human community (real and/or imagined) who are present and concerned. Experiences of exposure take place with an awareness of the presence of the others involved. From this, it is easy to recognize that embedded in any notion of risk is a community, physically embodied and/or bodily imagined.

While the challenge of exposure is to stand (unmasked/unmade-up/vulnerable) in the presence of others, the challenge of danger is harm (broadly defined). With the presence of danger is the likelihood of direct harm, and harm is what humans and human communities seek to avoid.

Risk points to the reality that processes of learning and performance are wrought with exposure and danger, with the known, the unknown, and un/knowable intermingled in unpredictable and, at times, unsettling ways. Learning and performance are risky businesses; both processes hold within them the potential for harm. AM5's brief reflection, cited above, captures the troubling relationship between performance (from an audience perspective) and community building, and points to the implicit risk involved in both. AM5 describes the experience of being an audience member in the *Break in' Point* performance as *distressing*, though it was a performance s/he "*really really enjoyed"*. AM5's reflection also captures the experience of an audience member who invested in being fully present during the performance, willingly exposed, s/he was responsive to and incorporated in the contact zone in which audience members and performers engaged in highly affective exchanges that ignited distress and joy. One of the paradoxes of embodied/somatic processes is that distress and joy, along with the spectrum of emotions between and beyond these two markers, are, at times, simultaneously intermingled and all-consuming. As a result, such processes can be unsettling; however, these unsettled spaces are fertile ground for deep learning. Audience members offered the following:

> "They related so well... and yet they were so different... I felt so consumed by it" AM17.
> "At times the lines between dancer and physicist just blurred. . . Wow!!!" AM 25
> "I feel I was so involved in it all from the very beginning when we were asked to dance... like all of me" AM33
> "I never thought I would understand anything having to do with physics, but in this 30 minutes I have learned more about physics and basic human resilience than I ever imagined" AM56.

Conscious consensual investment is required at varying levels in processes of learning and performance. Learning and performance, in the making of temporal communities, expose students, audience members and performers to possibilities for acquiring new knowledge and new ways-of-knowing that become incorporated into systems of living, being in and constructing meaningful ways to understand the world.

## Discussion

The creation of temporal communities allows participating members to claim belonging to and ownership of learning and performance processes. The sense of ownership potentially acts as a cushion to real or imagined danger embedded in learning and performance processes. The *Break-* *in' Point* performance and video offered opportunities to experience and witness varying levels of community building, articulations of belonging through exposure to vulnerabilities and collective narratives of resilience. The assent to be a part of Break-in' Point, a site of human experimentation, revealed the driving nature of curiosity - pointing to a willingness to take on a certain percentage of calculated and un-calculated risk and a hopefulness towards outcomes. Somatic work is intellectual, where the body is the articulated and articulating site of intelligence, making human experimentation experiential and risky. With the body as the primary access to knowledge, knowledge is emotional, sensorial; it is rich and thick with humanity. Audience member#92 notes:

> "From the moment the physicist [A2] fell... then [A3] fell but, he was a physicist falling and then the A3 [himself] fell . . . but everyone kept getting back up... they kept on... It was so emotional . . . so moving." (AM92).

When embodied experimentation is centered in the learning process, the potential for deep learning is ignited. Deep learning *is* realized. Affect and somatic ethics allow embodied community members to connect with one another and stay connected and emotionally vulnerable and responsive. Deep learning requires that individuals in communities "keep getting back up;" in the performance and the building of temporal communities, encounters and exposures with vulnerability are possible

In the section that follows, A4 addresses bodily knowledge and embodied learning, exploring encounters with *Break in' Point* in the virtual (imagined) somatic zone in processes of learning beyond the live theatre and within the conventional classroom.

## Beyond live theatre: Digital Engagement & Somatic Pedagogy - Meaning Making in the Classroom

Digital engagement offers the opportunity for live theatre to reach beyond initial target communities of theatre professionals and live public audiences. The spaces in which theatre professionals and live audiences engage are spaces rich with semiotics that informs and are formed by affect. But what of this is transmittable through digital medium and beyond the live theatre is difficult to discern. Common language and gesturing patterns that are often culturally bound can potentially place limits on the quality and/or breadth of experiences that are engaged through digital media. In an embodied cognition context the human body is a semiotic resource that fosters active learning in terms of cooperation with components that comprise the external environment. The Stanford Encyclopedia of Philosophy (http://plato.stanford.edu/entries/embodied-cognition/) online resource on embodied cognition notes that:

> "[…]many features of cognition are embodied in that they are deeply dependent upon characteristics of the physical body of an agent, such that the agent's beyond-the-brain body plays a significant causal role, or a physically constitutive role, in that agent's cognitive processing."

In the field of science education, a significant portion of researchers approach learning as a process of actions with the human body (gestures, displacements, inclination of the trunk) holding a dominant role. Bodily actions may represent concepts and endow with meaning non-visible or abstract entities such as movements of electrons or field lines (Roth, 2001), whilst also expressing dimensions, which verbal discourse cannot communicate.

Students who internalize sensorimotor activities at a preliminary rehearsal stage retain the ability to apply such somatic rules in future tasks (Hadzigeorgiou et al., 2009). In experimental teaching, research has demonstrated that in the mental processing of data the teacher's gestures precede or replace verbal discourse (Roth and Lawless, 2002). Moreover, gestures, change in body stance, movements in space, and actions involving material objects form conceptual links between what is being uttered by the teacher and the spatial components of the learning environment (Hwang and Roth, 2011). This section addresses the movements and gestures of the body in digital and live forms within an explicit teaching/learning environment.

## *Break-in' Point* Context

The digital video version of the *Break-in' Point* performance was used as educational material in the course *Semiotics of Science Teaching,* offered in the Faculty of Education at Aristotle University of Thessaloniki in Greece. In this course, students are trained to approach wordings, material objects and the human body as dynamic vehicles which serve equally in the construction of scientific concepts. In the use of the digital version of Break-in' Point, the teacher exploits the pedagogical value of digital media with the aim to investigate ways in which students - in their effort to understand physics concepts conveyed in the video - engage and conceptualize through their bodies.

This experiment included a teacher (A4) and three students in the context of the aforementioned course, delivered in seminar-type forum: Students and teacher viewed the video of the *Break-in' Point* performance and began the process of assigning verbal explanations whilst analyzing the actions of the performers, A2 and A3.

## Process

The process of the experiment included three primary activities: 1) watching the video, 2) identifying actions and gestures relevant to physics concepts, and 3) students (and teacher) physically engaging (or replicating) and translating the actions and gestures from the video with their own bodies, sensing and feeling knowledge through essential bodily actions, gestures, experiences and encounters. This experiment was recorded, ultimately providing another digital product and tool. The three students and the teacher (A4) watched the video of *Break-in' Point* together, as a small community. It was the students' first time viewing the video; each time a student or the teacher (A4) identified moments believed to be relevant (movements or gestures performed by A2 and A3) to the aim of understanding a relevant physics concept, the video was paused and reviewed. An important part of the experiment was that throughout the process, students were encouraged to stand up, move around the room and provide explanations using their bodies. Students experientially felt and recognized sensations as a means of learning, while both using the body as a vehicle to act out or replicate what they saw and constructing meaning through this process.

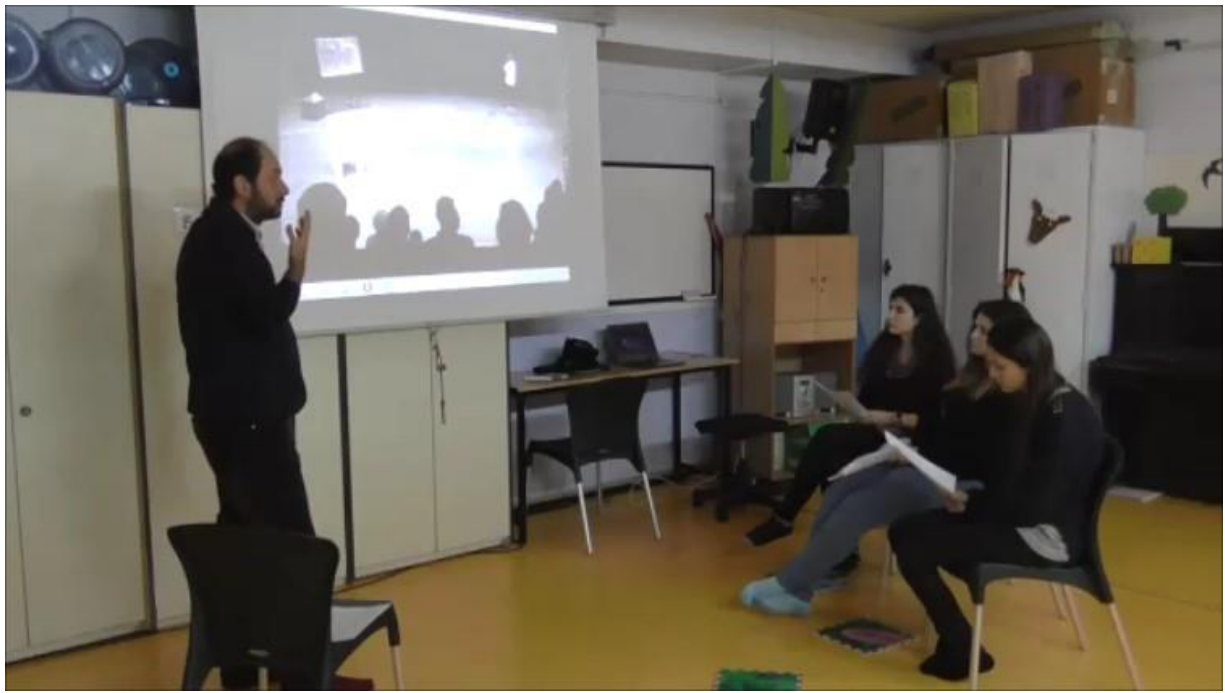

*Figure 6 A4 and students watch the Break-in' Point video during a lecture on education*
.

Examples of such moments include A2's explanation of the concept of force (Newton's second law) and A3's movements on the stage visualizing various types of motion (force, acceleration).

When students' comments and questions revealed important points of interest the video was paused and rewound or forwarded, in order to pay closer attention to the relationship and exchange that took place within the digitized space. A2 and A3's bodily-based explanations of physics concepts reflected the teacher and the students' understandings. Through activation of bodies, with gestures and poses, the students and teacher (A4) deepened understandings of what took place in the video, and were able to further access the *Break-in' Point* use of human vulnerability and resilience for illustrating physics concepts. Students also gained proprioceptive feedback of motion in space rather than just demonstrating the actions in space.

Additionally, *Break-in' Point* allowed for moments of *imagined* sharing between digital and live actors; moments that engendered unique types of temporal communities that transcend the particularities of the entities can be brought together creating real or imagined landscapes (zones) (Schwartz, 2006).

## Analysis

In the classroom, students and teacher (A4) generate actions for interpretation (active narrators on stage). Initially, the video of *Break-in' Point* was projected in the classroom. Subsequently, meaning was constructed by actions coming from the projected digitized performance space and the interactions within the physical space of the classroom. The digitized video is a pre-constructed non-three-dimensional space - created as a physical entity in the past and registered through its digital version in a storage medium. The video of *Break-in' Point* is a digitized space that transfers image information, in this instance, into the classroom setting. The video is inert in nature, neither the teacher nor the students can directly interact with it, change, or alter its course; with the digitized space such possibilities do not exist.  In contrast, the typical physical space, such as the classroom, is an active space that is valued, among other things, for its potential for change. In general, these kinds of spaces can be constructed as movable and non-movable and can be animate or/and inanimate. In this experiment, the physical space in Figure 6 is constructed of the walls, the chairs, some props, and the projection screen, and also of the persons contained in it whether still or moving. Central to this project is the changeable nature of the classroom that reinforced the assignment of meaning through human interaction.

In the meaning-making process three functions of (inter)actions between digitized and physical space were identified: (a) incoming information, (b) transferability and (c) blended integration.

(a) Incoming information

Students and teacher (A4) received the information from digitized space into the physical space of the classroom (see Figure 7).

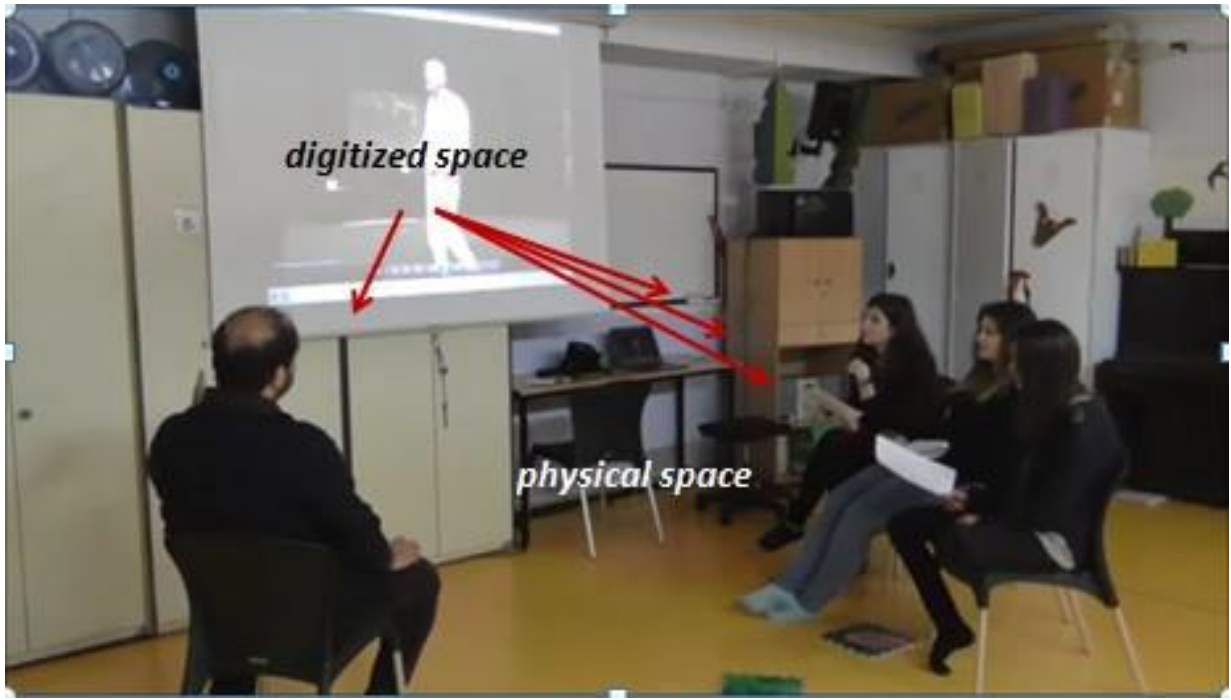

*Figure 7 A4 and student group watch passively the Break-in' Point video.*

(b) Transferability

Taking place in the physical space, in this instance in the classroom, transferability identifies bodily based interactions that apply elements or ideas from the digitized space of *Break-in' Point* to the physical space. In this stage, the students and A4 did not try to assign meaning explicitly to what took place in the digitized space. Instead, they took an idea from the video and, leaving the video's digitized space behind, produced only bodily interactions in the physical space. Simply executed, each time the projected video was paused, the students and teacher physically applied digital elements and worked to clarify the meaning/s on their own bodily actions. These bodily based interactions produced movement figures that expressed thoughts and ideas, often producing new approaches or questioning an idea.

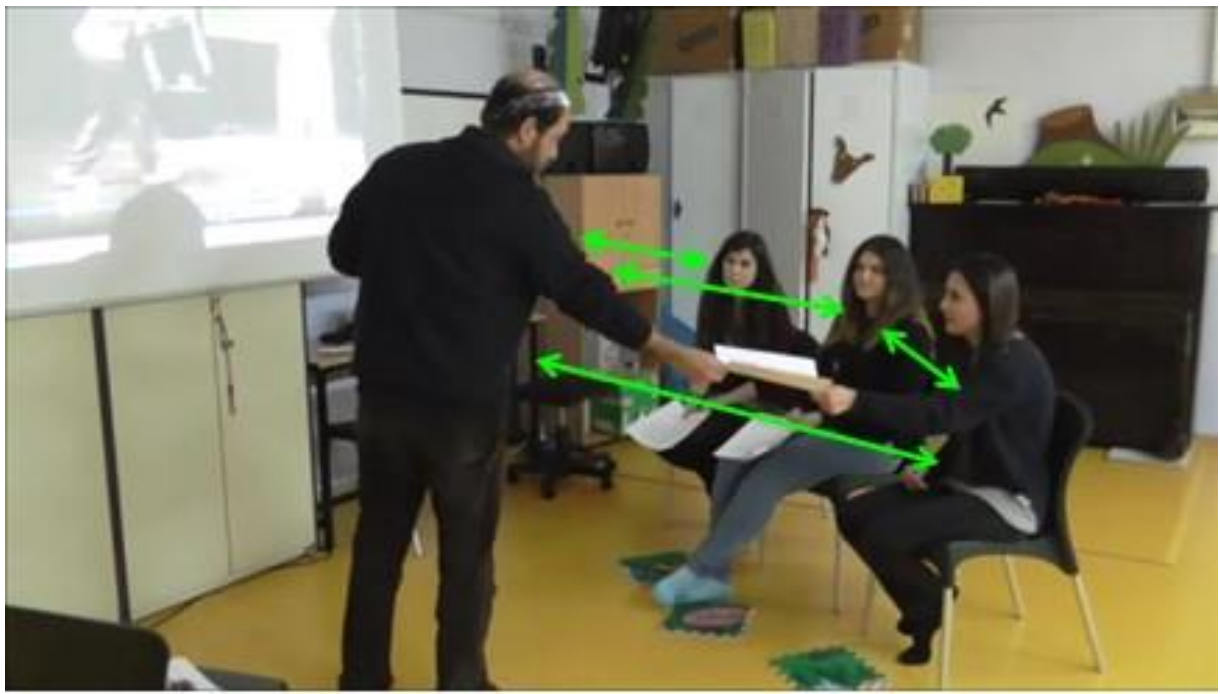

*Figure 8 A4 and students engage in somatic interactions in the physical space.*

In Figure 8 Student#1 and the teacher (A4) used a sheet of paper to introduce the concept of unequal forces. To understand mechanical equilibrium in terms of equal in value, but opposite forces – students took a sheet of paper, and directed by Student#1's proposal, the other students attempted to achieve equilibrium by both pulling the paper. Thus, the students and teacher (A4) grasped the idea of mechanical equilibrium introduced in the digitized space, transferred it to the physical space of the classroom producing a brand new bodily form (i.e., pulling the paper) and then assigned to it specific meaning. In another instance (Figure 9 below) Student#1 explained by means of her body why, when she is seated on a chair, she is not moving. In both cases, Student#1 indicated with her hands the direction of the exerted forces, producing two figures; one visualizing the direction of her weight (downwards) and the other performing the reaction force exerted on her from the chair (upwards).

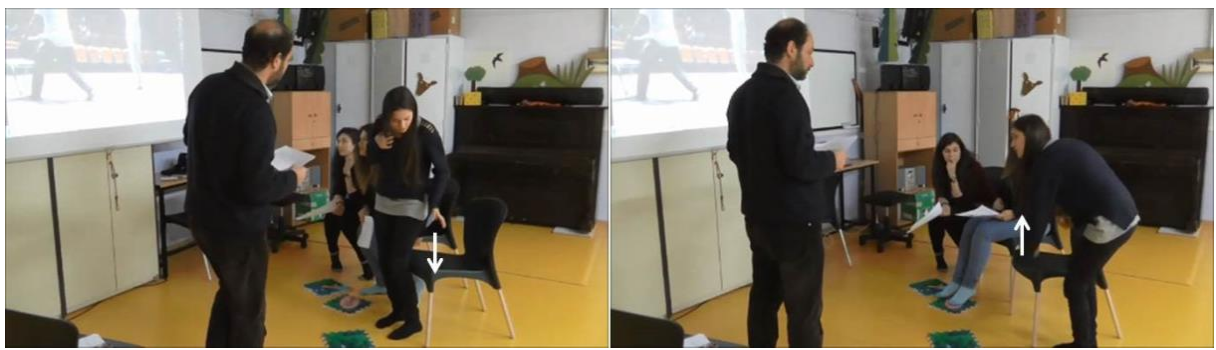

*Figure 9 In both cases, Student#1 moves her hand upwards and downwards indicating the two forces that are in balance.*

    (c) Blended integration

Blended integration refers to any combination of experience and knowledge available in or through the physical and the digitized spaces. In this instance, students worked to understand the actions in the *Break-in' Point* performance in terms of the somatic knowledge acquired in the experimentations with their own physical interactions in the classroom. In Figure 10, Student#2 and the teacher, in their effort to conceptualize a complex series of movement, moved in rotating formation through the classroom space (replicating A3 in the digitized space dancing holding a chair) and subsequently, explained the digital images in terms of somatic sensations produced in the physical space, ultimately creating new knowledge. This process is understood as a blended integration since meaning is constructed at the intersection of the two narrative spaces, the classroom setting and the projected video of *Break-in' Point*.

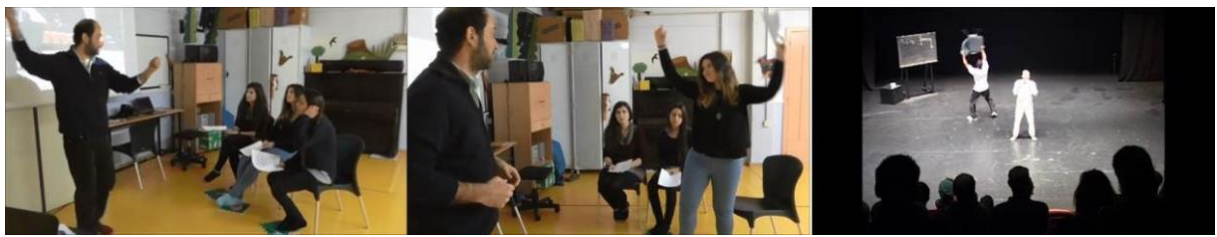

*Figure 10 A4's body starts to rotate and move forward. Student#1 performs a rotation through physical space and a complex motion is created and bodily experienced. A2 refers to A3's performance, A3, holding the chair performs complex motion on the stage.*

## Discussion

In the classroom setting, the students and teacher (A4) received the incoming information from digitized space with their bodies not directly activated. Gradually, this information was consciously transferred through bodily interactions between A4 and the students in the physical space where meaning was assigned. In some instances these interactions produced totally different figures when compared to A2 and A3's actions in the digitized space.

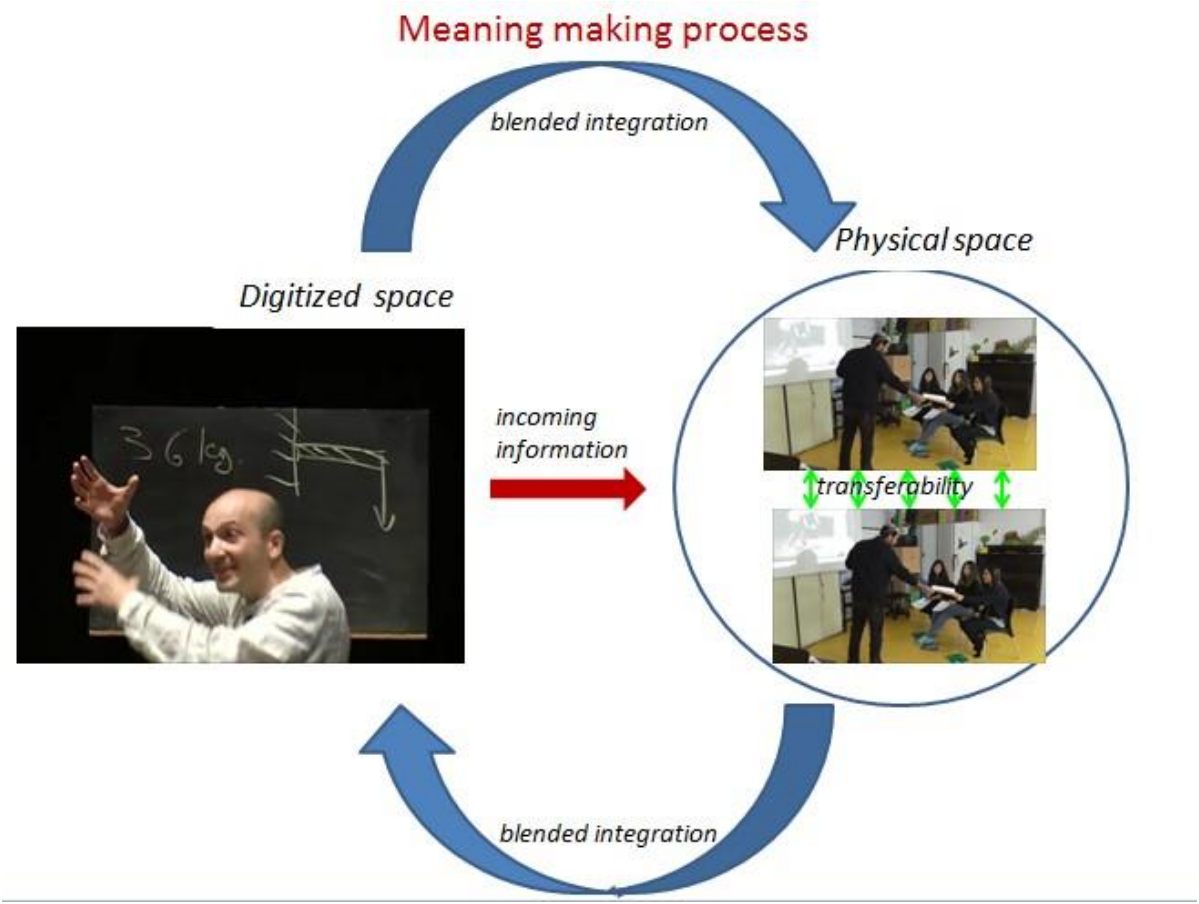

*Figure 11 Incoming information, transferability and blended integration functions of a meaning making process.*

Finally, a type of a blended integration took place in the third stage of meaning/making process. By again restructuring meaning that students and teacher (A4) had previously internalized into their collective bodies, new meaning was re-imagined in A2 and A3's actions in the digitized space.

## Summing Up

This article examined the life of *Break-in' Point,* reviewing and analyzing it as an imagined somatic zone that engaged spiritual and epistemological transformation of performers and audiences alike. We set out to build communities of healing and resilience by starting with ourselves and then shared the process with the communities around us. Even before the time of ancient Greek society, theatre had central educational value in the development of communities and social actors. And, from its inception, *Break-in' Point* aimed to address students at the university, the wider university community and the general public in conversations around human vulnerability and the intermingling of physics and dance in human life. Our endeavor was multifaceted, from concept to rehearsal to stage and audience to virtual and classroom reenactment, *Break-in' Point* offered organic ways and continuous opportunities to build diverse communities centered on human understanding of self and others.

There are several important lessons *we* have learned relevant to teaching and learning, trauma and resilience, and the building of community. In the beginning, we were technically strangers to each other; yet, over a relatively short but intense period of time, we emerged with a collaboration that deepened our understanding of ourselves and of teaching and learning processes in and outside classroom settings. Surprisingly, we witnessed important and unexpected transformations as the human form passed through various media: studio rehearsals, theatrical performance, video digital form, internet and the classroom. We came to realize that new technologies can facilitate better understanding of complex concepts (in the humanities and in science) and can be deployed to increase human interactions and strengthen the fabric of communities, or build new communities in unexpected ways. The virtual form of the project made possible the exposure of *Break-in' Point* to wider and more diverse communities, the propagation to interdisciplinary alliances, and its transformation to different concepts. Through the internet it was possible to understand the impact of *Break-in' Point* on teaching and learning at the university and in society, this impact expanded the initial concept of *Break-in' Point* beyond its own initial borders into a rich and unimagined future.

# Bibliography

AHMED, S. 2004. *The Cultural Politics of Emotion,* Edinburgh, Edinburgh University Press.

ASSAF, N. M. 2012. I Matter: an interactive exploration of audience–performer connections. *Research in Dance Education,* September**,** 1-27.

BREAK-IN' POINT(Part I) (2012) YouTube video, added by Jiannis Pachos [Online]. Available at https://www.youtube.com/watch?reload=9&v=wqBWcHUjLeU.

BREAK-IN' POINT (Part II) (2012) YouTube video, added by Jiannis Pachos [Online]. Available at https://www.youtube.com/watch?v=_o3K9pmMcXY.

BRENNAN, T. 2004. *The Transmission of Affect,* Ithaca and London, Cornell University Press.

DE CERTEAU, M. 1984. *The Practice of Everyday Life,* Berkeley/Los Angeles/London, University of California Press.

GEBER, P. & WILSON, M. 2010. Teaching at the Interface of Dance Science and Somatics *Journal of Dance Medicine & Science* 14**,** 50 - 57.

GEERTZ, C. 1973. *The Interpretation of Cultures,* New York Basic Books.

HADZIGEORGIOU, Y., Anastasiou, L., Konsolas, M., & Prevezanou, B. 2009. A study of the effect of preschool children's participation in sensorimotor activities on their understanding of the mechanical equilibrium of a balance beam. *Research in Science Education*, *39*(1), 39-55.

HWANG, S., & ROTH, W. M. 2011. The (embodied) performance of physics concepts in lectures. *Research in Science Education*, *41*(4), 461-477.

HARRISON, P. C., WALKER III, V. L. & EDWARDS, G. 2002. *Black Theatre: Ritual Performance in the African Diaspora,* Philadelphia, Temple University Press.

LEPECKI, A. 2006. P*erformance and the Politics of Movement*, New York and London: Routledge.

PROFETA, K. 2015. *Dramaturgy in Motion: At Work on Dance and Movement Performance,* Madison, WI and London, University of Wisconsin Press.

ROTH, W. M. 2001. Gestures: Their role in teaching and learning. *Review of educational research*, *71*(3), 365-392.

ROTH, W. M., & Lawless, D. 2002. Scientific investigations, metaphorical gestures, and the emergence of abstract scientific concepts. *Learning and instruction*, *12*(3), 285-304.

SCHWARTZ, L. 2006. Fantasy, realism, and the other in recent video games. *Space and Culture,* 9(3), 313–325.